\documentclass[epsf]{elsart}

\usepackage{graphicx}
\usepackage{amssymb}

\usepackage{epsfig} 
\usepackage{changebar} 
\usepackage{amsmath}

\journal{Physica A}

\newcommand{\be}{\begin{equation}}
\newcommand{\ee}{\end{equation}}
\newcommand{\bea}{\begin{eqnarray}}
\newcommand{\eea}{\end{eqnarray}}

\newcommand{\avg}[1]{\langle\langle{#1}\rangle\rangle}

\begin{document}

\begin{frontmatter}

\title{Theory of rumour spreading in complex social networks}
\author[bt]{M. Nekovee},
\author[bifi]{Y. Moreno},
\author[ictp]{G. Bianconi} and
\author[ictp]{M. Marsili}

\address[bt]{BT Research, Polaris 134, Adastral Park, Martlesham, Suffolk, IP5 3RE, UK}
\address[bifi]{Institute for Biocomputation and Physics of Complex
Systems and Department of Theoretical Physics, University of Zaragoza,
Zaragoza 50009, Spain}
\address[ictp]{The Abdus Salam International Centre for Theoretical
Physics, Strada Costiera 11, 34014 Trieste, Italy}

\begin{abstract}
We introduce a general stochastic model for the spread of rumours,
and derive mean-field equations that describe the dynamics of the
model on complex social networks (in particular those mediated by
the Internet). We use analytical and numerical solutions of these
equations to examine the threshold behavior and dynamics of the
model on several models of such networks: random graphs,
uncorrelated scale-free networks and scale-free networks with
assortative degree correlations. We show that in both homogeneous
networks and random graphs the model exhibits a critical threshold
in the rumour spreading rate below which a rumour cannot propagate
in the system. In the case of scale-free networks, on the other
hand, this threshold becomes vanishingly small in the limit of
infinite system size. We find that the initial rate at which a
rumour spreads is much higher in scale-free networks than in random
graphs, and that the rate at which the spreading proceeds on
scale-free networks is further increased when assortative degree
correlations are introduced.  The impact of degree correlations on
the final fraction of nodes that ever hears a rumour, however,
depends on the interplay between network topology and the rumour
spreading rate. Our results show that scale-free social networks are
prone to the spreading of rumours, just as they are to the spreading
of infections. They are relevant to the spreading dynamics of chain
emails, viral advertising and large-scale  information dissemination
algorithms on the Internet.
\end{abstract}

\end{frontmatter}

\section{Introduction}

Rumours are an important form of social communications, and their
spreading plays a significant role in a variety of human affairs.  The
spread of rumours can shape the public opinion in a country
\cite{galam}, greatly impact financial markets \cite{kimmel,kosfeld}
and cause panic in a society during wars and epidemics outbreaks.  The
information content of rumours can range from simple gossip to
advanced propaganda and marketing material. Rumour-like mechanisms
form the basis for the phenomena of viral marketing, where companies
exploit social networks of their customers on the Internet in order to
promote their products via the so-called `word-of-email' and
`word-of-web' \cite{viral1_domingos}.  Finally, rumor-mongering forms
the basis for an important class of communication protocols, called
gossip algorithms, which are used for large-scale information
dissemination on the Internet, and in peer-to-peer file sharing
applications \cite{gossip_demers,gossip2_review}.

Rumours can be viewed as an ``infection of the mind'', and their
spreading shows an interesting resemblance to that of epidemics.
However, unlike epidemic spreading quantitative models and
investigations of rumour spreading dynamics have been rather
limited. An standard model of rumour spreading, was introduced many
years ago by Daley and Kendall \cite{DK1,book1_epidemic}.
The Daley-Kendall (DK) model and its variants, such
as the Maki-Thompson (MK) model \cite{Maki}, have been used
extensively in the past for quantitative studies of rumour spreading
\cite{DK2,sudbury,pittel,noymer}.  In the DK model a closed and
homogeneously mixed population is subdivided into three groups, those
who are ignorant of the rumour, those who have heard it and actively
spread it, and those who have heard the rumour but have ceased to
spread it.  These groups are called ignorants, spreaders and stiflers,
respectively. The rumour is propagated through the population by {\em
pair-wise} contacts between spreaders and others in the population,
following the law of mass action. Any spreader involved in a pair-wise
meeting attempts to 'infect' the other individual with the rumour. In
the case this other individual is an ignorant, it becomes a
spreader. In the other two cases, either one or both of those involved
in the meeting learn that the rumour is `known' and decided not to
tell the rumour anymore, thereby turning into stiflers
\cite{book1_epidemic}.  In the Maki-Thompson variant of the above
model the rumour is spread by {\em directed} contacts of the spreaders
with others in the population. Furthermore, when a spreader contacts
another spreader only the {\em initiating} spreader becomes a
stifler.
 
An important shortcoming of the above class of models 
is that they either do not take into account
the topology of the underlying social interaction networks along which
rumours spread (by assuming a homogeneously mixing population), or use
highly simplified models of the topology \cite{sudbury,pittel}. While
such simple models may adequately describe the spreading process in
small-scale social networks, via the word-of-mouth, they become highly
inadequate when applied to the spreading of rumours in large social
interaction networks, in particular those which are mediated by the
Internet.  Such networks, which include email networks
\cite{email_newman,email_ebel,p2pfang}, social networking sites
\cite{soc_nets1} and instant messaging networks \cite{instant}
typically number in tens of thousands to millions of nodes. The
topology of such large social networks shows highly complex
connectivity patterns. In particular, they are often characterized by
a highly right-skewed degree distribution, implying the presence of a
statistically significant number of nodes in these networks with a
very large number of social connections
\cite{email_newman,email_ebel,soc_nets1,instant}.

A number of recent studies have shown that introducing the complex
topology of the social networks along which a rumour spreads can
greatly impact the dynamics of the above models.  Zanette performed
simulations of the deterministic Maki-Thompson model on both static
\cite{rumour1_zanette} and dynamic \cite{rumour2_zanette} small-world
networks.  His studies showed that on small-world networks with
varying network randomness the model exhibits a critical transition
between a regime where the rumour ``dies'' in a small neighborhood of
its origin, and a regime where it spreads over a finite fraction of
the whole population. Moreno {\it et al.} studied the stochastic
version of the MK model on scale-free networks, by means of Monte
Carlo simulations \cite{rumour1_maziar}, and numerical solution of a
set of mean-field equations \cite{rumour2_maziar}. These studies
revealed a complex interplay between the network topology and the
rules of the rumour model and highlighted the great impact of network
heterogeneity on the dynamics of rumour spreading. However, the scope
of these studies were limited to {\it uncorrelated} networks. An
important characteristic of social networks is the presence of
assortative degree correlations, i.e. the degrees of adjacent vertices
is positively correlated \cite{review_newman,
review_yamir,email_newman, email_ebel}. Furthermore the mean-field
equations used in \cite{rumour2_maziar} were postulated without a
derivation.

In this paper we make several contributions to the study of rumour
dynamics  on complex social networks. First of all, we introduce a
new model of rumour spreading on complex networks which, in
comparison with previous models, provides a more realistic
description of this process. Our model unifies the MK model of
rumour spreading  with the Susceptible-Infected-Removed (SIR) model
of epidemics, and has both of these models as it limiting cases.
Secondly, we describe a formulation of  this model on networks in
terms of  Interacting Markov Chains (IMC) \cite{imc1}, and use this
framework to derive, from first-principles, mean-field equations
for  the dynamics of rumour spreading on complex networks with
arbitrary degree correlations. Finally, we use approximate
analytical and exact numerical solutions of these equations to
examine both the steady-state and the time-dependent behavior of the
model on several models of social networks: homogeneous networks,
Erd\H os-R\'enyi (ER) random graphs, uncorrelated scale-free
networks and scale-free networks with assortative degree
correlations.

We find that, as a function of the rumour spreading rate, our
model shows a new critical behavior on networks with bounded degree
fluctuations, such as random graphs, and that this behavior is
absent in scale-free networks with unbounded fluctuations in node
degree distribution. Furthermore, the initial spreading rate of a
rumour is much higher on scale-free networks as compared to random
graphs. This spreading rate is further increased when assortative
degree correlations are introduced. The final fraction of nodes
which ever hears a rumour (we call this the final size of a rumour),
however, depends on an interplay between the model parameters and
degree-degree  correlations. Our findings are relevant to a number
of rumour-like processes taking place on complex social networks.
These include the spreading of chain emails and Internet hoaxes,
viral advertising and large-scale data dissemination in computer and
communication networks via the so-called gossip protocols
\cite{gossip_demers}.

The rest of this paper is organized as follows. In Section 2  we
describe our rumour model. In section 3  a  formulation of the
model within the framework of Interactive Markov Chains is given, and
the corresponding mean-field equations are derived. In section 4
we present analytical results for the steady-state behavior of our
model for both homogeneous  and inhomogeneous social networks.
This is followed in section 5
by numerical investigations of the steady-state and dynamics of the
model on several models of social networks: the ER random graph, the
uncorrelated scale-free networks and the assortatively
correlated SF networks. We close this paper in section 6 with conclusions.

\section{A general model for  rumour dynamics on social networks}

The spread of rumours is a complex socio-psychological process. An
adequate modeling of this process requires both a correct description
of the underlying social networks along which rumours spread and a
quantitative formulation of various behavioural mechanisms that
motivate individuals to participate in the spread of a rumour.  The
model described below is an attempt to formalize and simplify these
behavioral mechanisms in terms of a set of simple but plausible rules.

Our model is defined in the following way.  We consider a population
consisting of $N$ individuals which, with respect to the rumour, are
subdivided into ignorants, spreaders and stiflers.  Following Maki and
Thompson \cite{Maki}, we assume that the rumour spreads by {\it
directed} contact of the spreaders with others in the
population. However, these contacts can only take takes place along
the links of an undirected social interaction network $G=(V,E)$, where
$V$ and $E$ denote the vertices and the edges of the network,
respectively. The contacts between the spreaders and the rest of the
population are governed by the following set of rules
\begin{itemize}
\item
Whenever a spreader contacts an ignorant, the ignorant
becomes an spreader at a rate $\lambda$.
\item
When a spreader contacts another spreader or a stifler the
{\it initiating} spreader becomes a stifler at a rate  $\alpha$.
\end{itemize}
In the above, the first rule models the tendency of individuals to
accept a rumour only with a certain probability which, loosely
speaking, depends on the urgency or credibility of a rumour. The
second rule, on the other hand, models the tendency of individuals
to lose interest in spreading a rumour when they learn, through
contacts with others, that the rumour has become stale news, or is
false. In both the  Daley-Kendall and the Maki-Thompson rumour
models, and their variants, stifling is the only mechanism that
results in cessation of rumour spreading. In reality, however,
cessation can occur also purely as a result of spreaders forgetting
to tell the rumour, or their disinclination to spread the rumour
anymore. Following a suggestion in \cite{book1_epidemic}, we take
this important mechanism into account by assuming that individuals
may also cease spreading a rumour spontaneously (i.e. without the
need for any contact) at a rate $\delta$. The spreading process
starts with one (or more) element(s) becoming informed of a rumour
and terminates when no spreaders are left in the population.

%We note that a possible variant of the above model, which can be
%considered as an extension of the DK rumour model, is one in which
%both the spreading of the rumour and the stifling process result from
%{\it pairwise} contacts between all  members of the population, instead of
%directed contacts between spreaders and the rest.
%It can be  shown, however, that the
%deterministic mean-field equation for both variant of the model are
%identical \cite{book1_epidemic,maziar_unpub}. For this reason
%we shall limit ourselves in the rest of the paper to investigating
%only the first variant of our model.

\section{Interactive  Markov chain mean-field equations}
We can describe the dynamics of the above model on a network  within
the framework of the Interactive Markov Chains (IMC).  The IMC  was
originally introduced as a means for modelling social processes
involving many  interacting actors (or agents) \cite{imc1}. More
recently they have been applied to the dynamics of cascading
failures in electrical power networks \cite{imc2}, and the spread of
malicious software (malware) on the Internet \cite{imc3}. An IMC
consists of $N$ interacting nodes, with each node having a state
that evolves in time according to an internal Markov chain.  Unlike
conventional Markov chains, however, the corresponding internal
transition probabilities depend not only on the current state of a
node itself, but also on the states of all the nodes to which it is
connected.  The overall system evolves according to a global Markov
Chain whose state space dimension is the product of states
describing each node. When dealing with large networks, the
exponential growth in the state space renders direct numerical
solution of the IMCs extremely difficult, and one has to resort to
either Monte Carlo simulations or approximate solutions.  In the
case of rumour model, each internal Markov chain can be in one of
the three states: ignorant, spreader or stifler. For this case we
derive below a set of coupled rate equations which describe on a
mean-field level the dynamics of the IMC. We note that a similar
mean-field approach may also be applicable to other dynamical
processes on networks which can be described within the IMC
framework.

Consider now a node $j$
which is in the ignorant state at time $t$. We denote with $p^j_{ii}$
the probability that this node stays in the ignorant state in
the time interval $[t,t+\Delta t]$, and with
$p^j_{is}=1-p^j_{ii}$ the probability that it makes a transition to
the spreader state. It then follows that
\be
p_{ii}^{j}= (1-\Delta t \lambda)^{g},
\ee
where  $g=g(t)$ denotes the number of neighbors of node $j$
which are in the spreader state at time $t$. In order
to progress, we shall coarse-grain the micro-dynamics of the system by
classifying  nodes in our network according to their degree
and taking statistical average of the above transition probability
over degree classes.

Assuming that a node $j$ has $k$ links, $g$  can be considered
as an stochastic variable which has the following binomial distribution:
\be
\Pi(g,t)
=\binom{k}{g}\theta(k,t)^{g}(1-\theta(k,t))^{k-g},
\ee
where $\theta(k,t)$ is the probability at time $t$
that an edge emanating from
an ignorant  node with $k$ links points to a spreader node. This
quantity can be written as
\be
\theta(k,t)= \sum_{k'} P(k'|k)P(s_{k'}|i_k) \approx
\sum_{k'} P(k'|k)\rho^s(k',t) .
\ee
In this equation $P(k'|k)$ is
the degree-degree correlation function, $P(s_{k'}|i_k)$ the
conditional probability that a node with $k'$ links is in the spreader
state, given that it is connected to an ignorant node with degree $k$,
and $\rho^s(k',t)$ is the density of spreader nodes  at time
$t$ which belong to connectivity class $k$.
In the above equation the
final approximation is obtained by ignoring  dynamic correlations
between the states of neighboring  nodes.

The transition probability $\bar{p}_{ii}(k,t)$ averaged over all possible
values of $g$ is then given by:
\begin{eqnarray}
\bar{p}_{ii}(k,t) & = &
\sum_{g=0}^k \binom{k}{g}
(1-\lambda\Delta t )^g\theta(k,t)^{g}(1-\theta(k,t))^{k-g} \nonumber \\
& = &
\left( 1-\lambda \Delta t \sum_{k'}P(k'|k)\rho^s(k',t)\right)^k.
\end{eqnarray}

In a similar fashion we can derive an expression for the
probability
$\bar{p}_{ss}(k,t)$ that a spreader node which has
$k$ links stays in this state  in the interval
$[t,t+\Delta t]$. In this case, however, we also  need
to compute the expected value of the number of stifler neighbors
of the node at time $t$. Following steps  similar to the previous paragraphs
we obtain
\\
\begin{eqnarray}
\bar{p}_{ss}(k,t) &=&
\left( 1-\alpha \Delta t
\sum_{k'}P(k'|k)(\rho^s(k',t)+\rho^r(k',t))\right)^k \nonumber \\
&\times&  (1-\delta \Delta t).
\end{eqnarray}
The corresponding probability for a transition from the spreader
to the stifler state, $\bar{p}_{sr}(k,t)$ is given by $
\bar{p}_{sr}(k,t)=1-\bar{p}_{ss}(k,t)$.

The above transition probabilities can be used to set up a system of
coupled Chapman-Kolmogorov equations for the probability distributions
of the population of ignorants, spreaders and stiflers within each
connectivity class. However, ignoring fluctuations around expected
values we can also obtain a set of deterministic rate equations for
the expected values of these quantities. 

Denote with $I(k,t), S(k,t), R(k,t)$ the expected values of the 
populations of nodes
belonging to connectivity class $k$ which at time $t$ are in the ignorant,
spreader or stifler state, respectively.
The event that an ignorant  node in class $k$  
will make a transition to the spreader state during $[t,t+\Delta t]$
is a Bernoulli random variable with probability 
$(1-p_{ii}(k,t))$ of success. As a sum of i.i.d random Bernoulli
variables, the total number of successful transitions in this time
interval has a binomial distribution, with an expected value given by  
$I(k,t)(1-p_{ii}(k,t))$. Hence the rate of change in the expected value 
of the population of ignorant nodes 
belonging to class $k$ is given by
\begin{eqnarray}
I(k,t+\Delta t) &=&  I(k,t)-I(k,t)\nonumber \\
&\times & \left[1-\left(1-\lambda \Delta t \sum_{k'}\rho^s(k',t)P(k'|k)
\right)^k \right] 
\end{eqnarray}
Similarly, we can write the corresponding 
rate of  change in the population of spreaders and
stiflers as follows
\begin{eqnarray}
S(k,t+\Delta t) &=&S(k,t) + I(k,t)\left
[1-
\left(1-\lambda \Delta t \sum_{k'}\rho^s(k',t)P(k'|k)\right)^k\right]
\nonumber \\
&-&
S(k,t)\left[1-
\left(1-\alpha \Delta t
\sum_{k'}(\rho^s(k',t)+\rho^r(k',t))P(k'|k)\right)
^k\right] \nonumber \\
&-&
\delta S(k,t)
\end{eqnarray}
\begin{eqnarray}
R(k,t+\Delta t)&=&R(k,t) \nonumber \\
&+&
S(k,t) 
\left[1-
\left(1-\alpha \Delta t
\sum_{k'}(\rho^s(k',t)+\rho^r(k',t))P(k'|k)\right)^k\right]
\nonumber \\
&+&
\delta S(k,t)
\end{eqnarray}
In the above equations $\rho^i(k,t),\rho^s(k,t)$, and $\rho^r(k,t)$
are the fraction of nodes belonging to class $k$ which are in the 
ignorant, spreader and stifler states, respectively.
These quantities  satisfy the normalization
condition $\rho^{i}(k,t)+\rho^s(k,t)+\rho^r(k,t)=1$. 
In the limit $\Delta t \rightarrow 0$ we obtain: 
\be \frac{\partial
\rho^i(k,t)}{\partial t}=-k \lambda
\rho^i(k,t)\sum_{k'}\rho^s(k',t)P(k'|k)
\label{ceq1}
\ee
\begin{eqnarray}
\frac{\partial \rho^s(k,t)}{\partial t} &=&
k \lambda \rho^i(k,t) \sum_{k'}\rho^s(k',t)P(k'|k)
\nonumber \\
&-&
k\alpha \rho^s(k,t)
\sum_{k'}(\rho^s(k',t)+\rho^r(k',t))P(k'|k) - \delta \rho^s(k,t).
\label{ceq2}
\end{eqnarray}
\begin{eqnarray}
\frac{\partial \rho^r(k,t)}{\partial t} &=&
k \alpha \rho^s(k,t)
\sum_{k'}(\rho^s(k',t)+\rho^r(k',t))P(k'|k) + \delta \rho^s(k,t).
\label{ceq3}
\end{eqnarray}

For future reference we note here that  information on the underlying network
is incorporated in the above equations solely via the degree-degree
correlation function. Thus in our analytical and numerical
studies reported in the next section we do not need to generate any
actual network. All that is required is either an analytical expression
for $P(k'|k)$ or a numerical representation of  this quantity.

\section{Analysis}

\subsection{Homogeneous networks}
In order to understand some  basic features of our rumour model we first
consider the case of homogeneous networks, in which degree
fluctuations are very small and there are no degree correlations.
In this case the rumour equations become:
\begin{eqnarray}
\frac{d\rho^i(t)}{dt} & = &-\lambda \bar{k}\rho^i(t)\rho^s(t) \\
\frac {d\rho^s(t)}{dt} & = & \lambda \bar{k}\rho^i(t)\rho^s(t)
-\alpha \bar{k} \rho^s(t)(\rho^s(t)+\rho^r(t)) \nonumber \\
& - & \delta \rho^s(t)
\\
\frac{d\rho^r(t)}{dt} & =& \alpha\bar{k}\rho^s(t)(\rho^s(t)+
\rho^r(t)) +\delta \rho^s(t),
\end{eqnarray}
where $\bar{k}$ denotes the constant degree distribution of the network (or the
average value for networks in which the probability of finding a node with a
different connectivity decays exponentially fast).

The above system of equations can be integrated analytically using an
standard approach. In the infinite time limit, when
$\rho^s(\infty)=0$, we obtain the following transcendal equation for
$R=\rho^r(\infty)$, the final fraction of nodes which ever hear the
rumour (we call this the final rumour size) \be R=1-e^{ -\epsilon R}
\label{threshold-hom} \ee where \be \epsilon=
\frac{(\alpha+\lambda)\bar{k}}{\delta+\alpha\bar{k}}. \ee Eq.
(\ref{threshold-hom})  admits a non-zero solution  only if
$\epsilon>1$. For $\delta \neq 0$ this condition is fulfilled
provided \be \frac{\lambda}{\delta}\bar{k} >1 , \ee which is
precisely the same threshold condition as found in the SIR model of
epidemic spreading on homogeneous networks
\cite{sir1_yamir,sir1_ml}. On the other hand, in the special case
$\delta=0$ (i.e when the forgetting mechanism is absent)
$\epsilon=1+\lambda/\alpha>1$, and  so Eq. (14) always admits a
non-zero solution, in agreement with the result in
\cite{rumour2_maziar}.

The above result shows, however, that the presence of a forgetting
mechanism results in the appearance of a finite threshold in the
rumour spreading rate below which rumours cannot spread in homogeneous
networks. Furthermore, the value of the threshold is {\em independent}
of $\alpha$ (i.e. the stifling mechanism), and is the same as that for
the SIR model of epidemics on such networks.  This result can be
understood by noting that in the above equations the terms
corresponding to the stifling mechanism are of second order in
$\rho^s$, while the terms corresponding to the forgetting mechanism
are only of first order in this quantity. Thus in the initial state of
the spreading process, where $\rho^s\approx 0$ and $\rho^r\approx0$,
the effect of stifling is negligible relative to that of forgetting,
and the dynamics of the model reduces to that of the SIR model.

\subsection{Inhomogeneous networks}
Next we consider uncorrelated inhomogeneous networks. In such networks
the degree-degree correlations can be written as
\cite{book_vespignani}:

\be
P(k'|k)= q(k')=\frac{k'P(k')}{\langle k \rangle},
\label{uncorr}
\ee

where $P(k')$ is the degree distribution and $\langle k \rangle $ is
the average degree. In this case the dynamic of rumour spreading is
describe by Eqs. (\ref{ceq1}-\ref{ceq3}).  Eq. (\ref{ceq1}) can be
integrated exactly to yield:

\be
\rho^i(k,t)=\rho^i(k,0) e^{-\lambda k\phi(t)},
\ee
where $\rho^i(k,0)$ is the initial density of  ignorant nodes
with connectivity $k$, and we have introduced the auxiliary function
\be
\phi(t)=\sum_k q(k) \int_0^t
\rho^s(k,t')dt'\equiv \int_0^t \avg{\rho^s(k,t')}dt'.
\ee
In the above equation and hereafter we  use the shorthand notation
\be
\avg{O(k)}=\sum_k q(k) O(k)
\ee
with
\be
q(k)=\frac{kP(k)}{\langle k \rangle}.
\ee
In order to obtain an expression for the final size of the rumour, $R$,
it is more convenient to work with $\phi$. Assuming an homogeneous
initial distribution of ignorants, $\rho^i(k,0)=\rho^i_0$,
we can obtain a differential equation for this quantity by multiplying
Eq. (\ref{ceq2}) with $q(k)$ and summing over $k$. This yields after some
elementary manipulations:
\begin{eqnarray}
\frac{d\phi}{dt} =  1-\avg{e^{-\lambda k\phi}}) -
\delta \phi -  \alpha \int_0^t
\left[1-\avg{e^{-\lambda k\phi(t')}}\right] \avg{k\rho^s(k,t')}dt',
\label{phi2}
\end{eqnarray}
where, without loss of generality, we have also put $\rho^i_0\approx
1$.

In the limit $t\rightarrow \infty$  we have
$\frac{d\phi}{dt}=0$, and Eq. (\ref{phi2}) becomes:
\begin{eqnarray}
0  =  1-\avg{e^{-\lambda k\phi_\infty}}
- \delta\phi_\infty -\alpha\int_0^\infty
\left[1-\avg{e^{-\lambda k\phi(t')}}\right] \avg{k\rho^s(k,t')}dt',
\nonumber \\
\label{phiinf}
\end{eqnarray}
where $\phi_\infty=\lim_{t\rightarrow \infty} \phi(t)$.

For $\alpha=0$ Eq. (\ref{phiinf})  can be solved explicitly to
obtain $\Phi_{\infty}$ \cite{sir1_yamir}. For $\alpha\neq0$ we solve
(\ref{phiinf})  to leading order in $\alpha$. Integrating Eq.
(\ref{ceq2}) to zero order in $\alpha$ we obtain \be
\rho^s(k,t)=1-e^{-\lambda k\phi}-\delta \int_0^t e^{\delta(t-t')}
\left[1-e^{-\lambda k\phi(t')}\right]dt'+O(\alpha). \ee Close to the
critical threshold both $\phi(t)$ and $\phi_\infty$ are small.
Writing $\phi(t)=\phi_\infty f(t)$, where $f(t)$ is a finite
function, and working to leading order in $\phi_\infty$, we obtain
\be \rho^s(k,t)\simeq -\delta \lambda k \phi_\infty \int_0^t
e^{\delta(t-t')}f(t') dt'+O(\phi_\infty^2)+O(\alpha) \ee Inserting
this in Eq. (\ref{phiinf})  and expanding the exponential to the
relevant order in $\phi_\infty$ we find
\begin{eqnarray}
0  = \phi_\infty\left[\lambda\avg{k}-\delta-
\lambda^2\avg{k^2}(1/2+\alpha\avg{k}I)\phi_\infty\right]
+ O(\alpha^2)+O(\phi_\infty^3)
\end{eqnarray}
where $I$ is a finite and positive-defined  integral.
The non-trivial solution of this
equation is  given by:
\be
\phi_\infty=\frac{\lambda\avg{k}-\delta}
{\lambda^2\avg{k^2}(\frac{1}{2}+\alpha I\avg{k})}.
\ee
Noting that  $\avg{k}=\langle k^2 \rangle/\langle k \rangle$
and $\avg{k^2}=\langle k^3 \rangle /\langle k \rangle $
we obtain:
\be
\phi_\infty=\frac{ 2\langle k \rangle
( \frac{\langle k^2 \rangle}{\langle k \rangle}\lambda -\delta) }
{\lambda^2 \langle k^3 \rangle (1+ 2\alpha I
\frac{ \langle k^2 \rangle}{\langle k \rangle})}.
\label{phi}
\ee
This yields a positive value for $\phi_{\infty}$ provided that
\be
\frac{\lambda}{\delta} \geq \frac{\langle k \rangle}{\langle k^2 \rangle}.
\label{threshold}
\ee
Thus, to leading order in $\alpha$, the critical
rumour threshold is independent of the stifling mechanism
and is the same as for the SIR model \cite{sir1_yamir,sir1_ml}.
In particular, for $\delta=1$ the critical rumour
spreading threshold is given by
$\lambda_c= \langle k \rangle/\langle k^2 \rangle$, and
Eq. (\ref{phi}) simplifies to:
\be
\phi_\infty=\frac{ 2\langle k \rangle (\lambda -\lambda_c)}
{\lambda^2 \langle k^3 \rangle(\lambda_c+2\alpha I)}.
\ee

Finally, $R$ is given by 

\be 
R=\sum_kP(k)(1-e^{-\lambda k\phi_\infty}),
\label{exp} 
\ee 

The solution to the above equation depends on the form of $P(k)$. In
particular, for homogeneous networks where all the moments of the
degree distribution are bounded, we can expand the exponential in
Eq. (\ref{exp}) to obtain

\begin{eqnarray}
R & \approx & \sum_k P(k) \lambda k \phi_\infty =
\frac{ 2\langle k \rangle^2 (\lambda -\lambda_c)}
{\lambda \langle k^3 \rangle(\lambda_c+2\alpha I)},
\end{eqnarray}

which shows that $R \sim (\lambda-\lambda_c)$ in the
vicinity of the rumour threshold. For heterogeneous networks, one must
solve the equation for $P(k)$. This can be done for example, as for
the SIR model \cite{sir1_yamir}.

\section{Numerical results}

\subsection{Random graphs and uncorrelated scale-free networks}
We consider first {\it uncorrelated}
networks, for which the degree-degree correlations are given by
Eq. (\ref{uncorr} ). We shall consider  two classes of such networks.
The first class is the Erd\H os-R\'enyi random  networks, which
for large $N$
have a Poisson degree distribution:
\be
P(k)=e^{-k}\frac{\langle k \rangle ^k}{k!}.
\ee
The above degree distribution peaks at an average value $\langle k \rangle$
and shows small fluctuations around this value. The second class we consider
are scale-free networks which have a power law degree distribution:
 \be
P(k)=
\begin{cases}
Ak^{-\gamma}
& k_{\text{min}}\leq k
\\
0 & \text{otherwise.}
\end{cases}
\ee
In the above equation $k_{\text{min}}$
is the minimum degree of the networks and $A$
is a normalization constant. Recent studies of social networks on the
Internet indicates that many of these networks show highly
right-skewed degree distributions, which could often be modelled by a
power-law degree distribution \cite{email_ebel,soc_nets1,instant}.
For $2\leq \gamma \leq 3$ the variance of the above degree
distribution becomes infinite in the limit of infinite system size
while the average degree distribution remains finite.
We shall consider henceforth SF networks with $\gamma=3$.

Our studies of uncorrelated networks were performed using the above
forms of $P(k)$ to represent ER and SF networks, respectively.  The
size of the networks considered was $N=10^5$ and $N=10^6$, and the
average degree was fixed at $ \langle k \rangle =7$. For each network
considered we generated a sequence of $N$ random integers distributed
according to its degree distribution. The coupled set of differential
equation (\ref{ceq1}-\ref{ceq3}) were then solved numerically using an
standard finite difference scheme, and numerical convergence with
respect to the step size was checked numerically.  In the following
and throughout the paper all calculations reported are performed by
starting the rumour from a randomly chosen initial spreader, and
averaging the results over 300 runs with different initial spreaders.
The calculations reported below were performed for networks consisting
of $N=10^6$ nodes.

In our first set of calculations we set $\delta=1$ and investigate
the dynamics as a function of the rumour spreading rate $\lambda$
and the stifling rate $\alpha$.
First we focus on the impact of network topology
on  the final size of a rumour, $R$, which for inhomogeneous networks
is obtained from
\be
R=\sum_k\rho^r(k,t_{\infty}),
\ee
where $t_\infty$ denotes a sufficiently long time at which
the spreading process has reached  its steady state
(i.e. no spreaders are left in the network). In Fig. 1 $R$
corresponding to the ER network is
plotted as a function of $\lambda$, and for
several different values of the stifling parameter $\alpha$.
It can be seen that in this network $R$  exhibits a critical
threshold $\lambda_c$ below which a rumour cannot spread in the
network. Furthermore, just as in the case of homogeneous networks,
the value of the threshold does not depend on $\alpha$, and is
at $\lambda_c=0.12507$. This value is in excellent agreement
with the analytical results obtained in the previous section.
We also verified numerically that, in agreement with our
analytical findings, the behaviour of  $R$ in the vicinity of the critical
point can, be described with the form
\be
R \sim A(\lambda-\lambda_c),
\ee
where $A=A(\alpha)$ is a smooth and monotonically decreasing
function of $\alpha$. The results are  shown in Fig. 2 where $R$ is plotted as
function of $\lambda$ for a range of values of $\alpha$, together
with the corresponding fits.

Next we turn to our results for the SF network.
In Fig. 3 results for $R$  in this  network are shown. In
this case we also observe the presence of an $\alpha$-independent
rumour threshold,
albeit for much smaller spreading rates than for the ER network.
We have verified  numerically that in this case the threshold is
approached with zero slope, as can also be gleaned from Fig. 3.
Since the value of the threshold is independent of $\alpha$,
we can use the well-known result
for the SIR model (the $\alpha=0$ case) to conclude that in the
limit of infinite system size the threshold seen in the
SF network will approach zero. It is therefore not an
intrinsic property of rumour spreading on this network.

In order to further analyze the behavior of $R$ in SF networks, we
have numerically fitted our results to the stretched exponential
form, \be R \sim \exp(-C/\lambda), \ee with $C$ depending only
weakly on $\alpha$. This form was found to describe the epidemic
prevalence in both the SIS and the SIR model of epidemics
\cite{sis1_vesp,sir1_yamir}.  The results are displayed in Fig. 4,
and they clearly indicate that the stretched exponential form also
nicely describes the behavior of $R$ in our rumour model. This
result provides further support for our conjecture that the general
rumour model does not exhibit any threshold behavior on SF networks
(at least in the limit of infinite systems size).

In addition to investigating the impact of network topology on the
steady-state properties of the model, it is of great interest to
understand how the time-dependent behavior of the model is affected by
topology.  In Figs. 5 and 6 we display, as representative examples,
time evolution of, respectively, the total fractions of stiflers and
spreaders, in both networks for $\lambda=1$ and two sets of values of
the cessation parameters: $\{\delta=1,\alpha=0 \}$, and $\{\delta=0,
\alpha=1\}$. The first set of parameters corresponds to a spreading
process in which cessation results purely from spontaneous forgetting
of a rumour by spreaders, or their disinclination to spread the rumour
any further.  The second set corresponds to a scenario where
individuals keep spreading the rumour until they become stiflers due
to their contacts with other spreaders or stiflers in the network.  As
can be seen in Fig. 5, in the first scenario the initial spreading
rate of a rumour on the SF network is much faster than on the ER
network. In fact, we find that the time required for the rumour to
reach $50\%$ of nodes in the ER random graph is nearly twice as long
as the corresponding time on the SF networks.  This large difference
in the spreading rate is due to the presence of highly connected nodes
(social hubs) in the SF network, whose presence greatly speeds up the
spreading of a rumour.  We note that in this scenario not only a
rumour spreads initially faster on SF networks, but it also reaches a
higher fraction of nodes at the end of the spreading process.

It can be seen from Figs, 5 and 6 that in the second spreading scenario
(i.e. when stifling is the only mechanism for cessation) the
initial spreading rate on the SF network is, once again, higher than
on the ER network. However, unlike the previous situation, the ultimate
size of the rumour is higher on the ER network.
This behavior is due to the conflicting
roles that hubs play when the stifling mechanism is switched
on. Initially the presence of hubs speeds up the spreading but once
they turn into stiflers they also effectively impede further spreading of the
rumour.

\subsection{Assortatively correlated scale-free networks}
Recent studies have revealed that social networks display
assortative degree  correlations, implying that highly connected
vertices preferably connect to vertices which are also highly
connected \cite{review_newman}.  In order to study the impact of
such correlations on the dynamics of our model, we make use of the
following ansatz for the degree-degree  correlation function
\cite{vazquez1}

\be
P(k'|k) = (1-\beta)q(k')+\beta \delta_{kk'}; \; \; \; \; \; \; \;
(0\leq\beta<1).
\ee
The above  form allows us to study
in a controlled way the impact of
degree correlations on the spreading of rumour.

Using the above degree-degree correlation function we numerically
solved Eqs. (\ref{ceq1}-\ref{ceq3}) for a SF network characterized by
$\gamma=3$ and $<k>=7$. The network size was fixed at $N=100,000$, and
we used two values for the correlation parameter: $\beta=0.2$ and
$\beta=0.4$. Fig. 7 displays $R$ as a function of $\lambda$, and for
$\alpha=0.5,0.75,1$ (the value of $\delta$ was fixed at $1$).

It can be seen that below $\lambda\approx 0.5$ a rumour will reach a
somewhat smaller fraction of nodes on the correlated networks than on
the uncorrelated ones.  However for larger values of $\lambda$ this
behavior reverses, and the final size of the rumour in assortatively
correlated networks shows a higher value than in the uncorrelated
network. We thus conclude that the qualitative impact of degree
correlations on the final size of a rumour depends very much on the
rumour spreading rate.

Finally, we investigate the effect of assortative correlations on the
speed of rumour spreading. In Fig. 8 we show our results for the time
evolution of the total fraction of spreaders, $S(t)$, in scale-free
networks consisting of $N=100,000$ nodes and for a correlation
strength ranging from $\beta=0$ to $\beta=0.4$.  In these calculations
the value of $\lambda$ was fixed at $1$, and we considered two values
of $\alpha$: 0,1. It can be seen that the initial rate at which a
rumour spreads increases with an increase in the strength of
assortative correlations regardless of the value of $\alpha$.
However, for $\alpha=1$ the rumour also dies out faster when such
correlations are stronger.

\section{Conclusions}

In this paper we introduced a general model of rumour spreading on
complex networks. Unlike previous rumour models, our model
incorporates two distinct mechanisms that cause cessation of a rumour,
stifling and forgetting. We used an Interactive Markov Chain
formulation of the model to derive deterministic mean-field equations
for the dynamics of the model on complex networks. Using these
equations, we investigated analytically and numerically the behavior
of the model on Erd\H os-R\'enyi random graphs and scale-free networks
with exponent $\gamma=3$. The critical behavior, the dynamics and the
stationary state of our model on these networks are significantly
different from the results obtained for the dynamics of simpler rumour
models on complex networks
\cite{rumour1_zanette,rumour2_zanette,rumour1_maziar,rumour2_maziar}.
In particular, our results show the presence of a critical threshold
in the rumour spreading rate below which a rumour cannot spread in ER
networks.  The value of this threshold was found to be independent of
the stifling mechanism, and to be the same as the critical infection
rate of the SIR epidemic model.  Such a threshold is also present in
the finite-size SF networks we studied, albeit at a much smaller
value. However in SF networks this threshold is reached with a zero
slope and its value becomes vanishingly small in the limit of infinite
network size.  We also found the initial rate of spreading of a rumour
to be much higher on scale-free networks than on ER random graphs. An
effect which is caused by the presence of hubs in these networks,
which efficiently disseminate a rumour once they become informed.  Our
results show that SF networks are prone to the spreading of rumours,
just as they are to the spreading of infections.

Finally, we used a local ansatz for the degree-degree correlation
function in order to numerically investigate the impact of assortative
degree correlations on the speed of rumour spreading on SF networks.
These correlations were found to speed up the initial rate of
spreading in SF networks.  However, their impact on the final fraction
of nodes which hear a rumour depends very much on the rate of rumour
spreading.

In the present work we assumed the underlying network to be static,
i.e. a time-independent network topology.  In reality, however, many
social and communication networks are highly dynamic. An example of
such time-dependent social networks is Internet chatrooms, where
individuals continuously make new social contacts and break old
ones. Modelling spreading processes on such dynamic networks is highly
challenging, in particular when the time scale at which network
topology changes becomes comparable with the time scale of the process
dynamics. We aim to tackle this highly interesting problem in future
work.

M.~N. acknowledges the Abdus Salam International Centre for
Theoretical Physics (ICTP) for a visiting fellowship during which some
amendments to this work were made. Y. M thanks the support of DGA
through a mobility grant (MI06/2005) and BT for hospitality. Y.~ M.
is supported by MEC through the Ram\'{o}n y Cajal Program. This work
was supported by BT and the Spanish DGICYT Projects
FIS2004-05073-C04-01 and FIS2005-00337.

%We thank Keith Briggs for reading the manuscript.
%\bibliographystyle{alpha}
%\bibliography{networks}% Produces the bibliography via BibTeX.

\newpage

\begin{figure}
 \epsfig{file=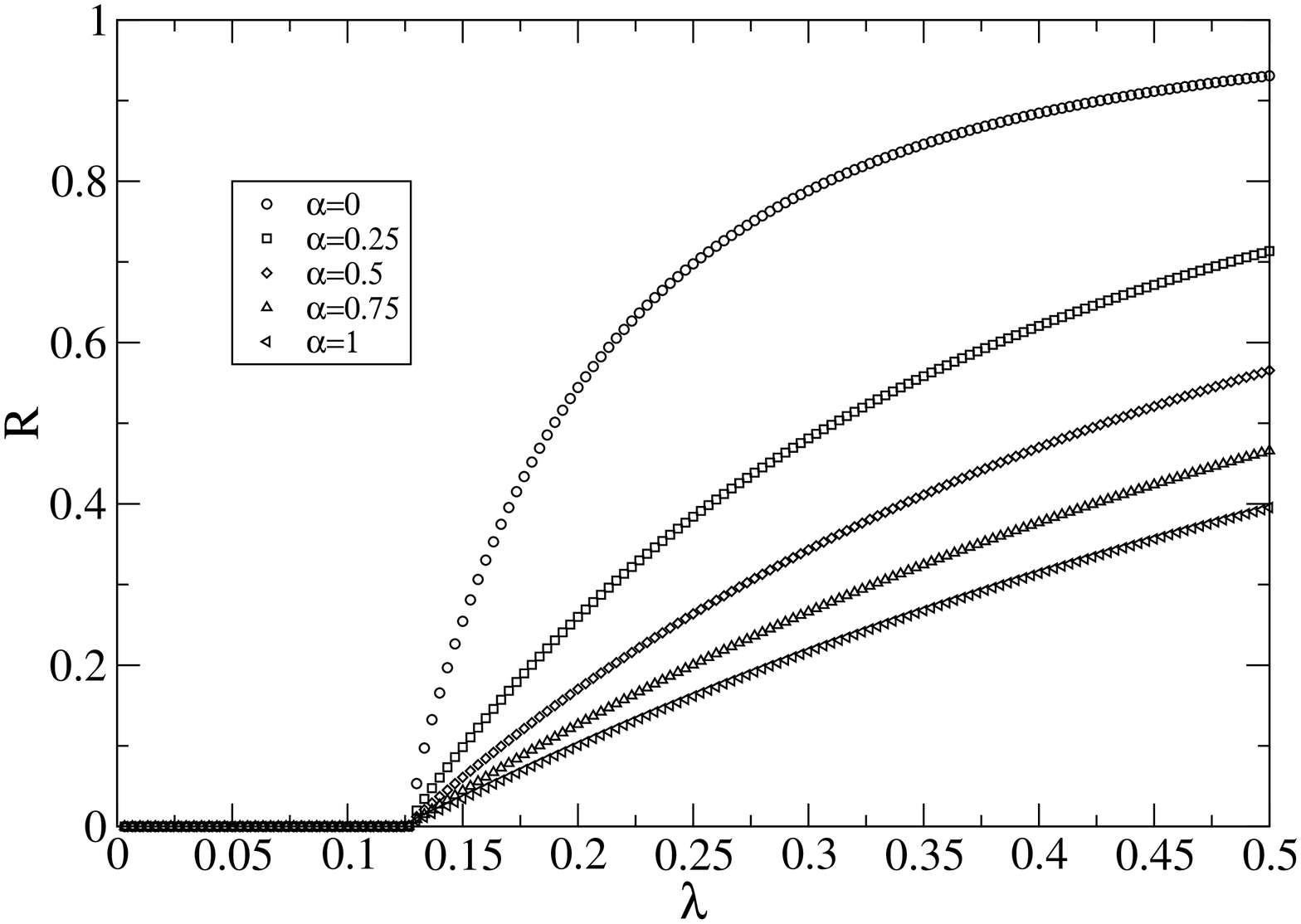,width=2.5in, angle=-90, clip=1}
\caption{ The final size of the rumor, $R$ is shown
as a function of the spreading rate $\lambda$
for the ER network of size $10^6$. The results are shown
for several values of the stifling parameter $\alpha$.}
\end{figure}

\begin{figure}
 \epsfig{file=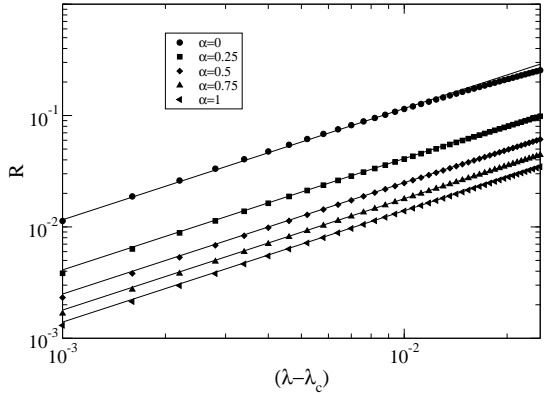,width=2.5in, angle=-90, clip=1}
 \caption{$R$ is plotted as a function of $\lambda-\lambda_c$ for the
ER network of size $10^6$, using different values of $\alpha$.
Solid lines show our numerical fits to the
form $R\sim (\lambda-\lambda_c)^\beta$, with $\beta=1$.}
\end{figure}

\begin{figure}
\epsfig{file=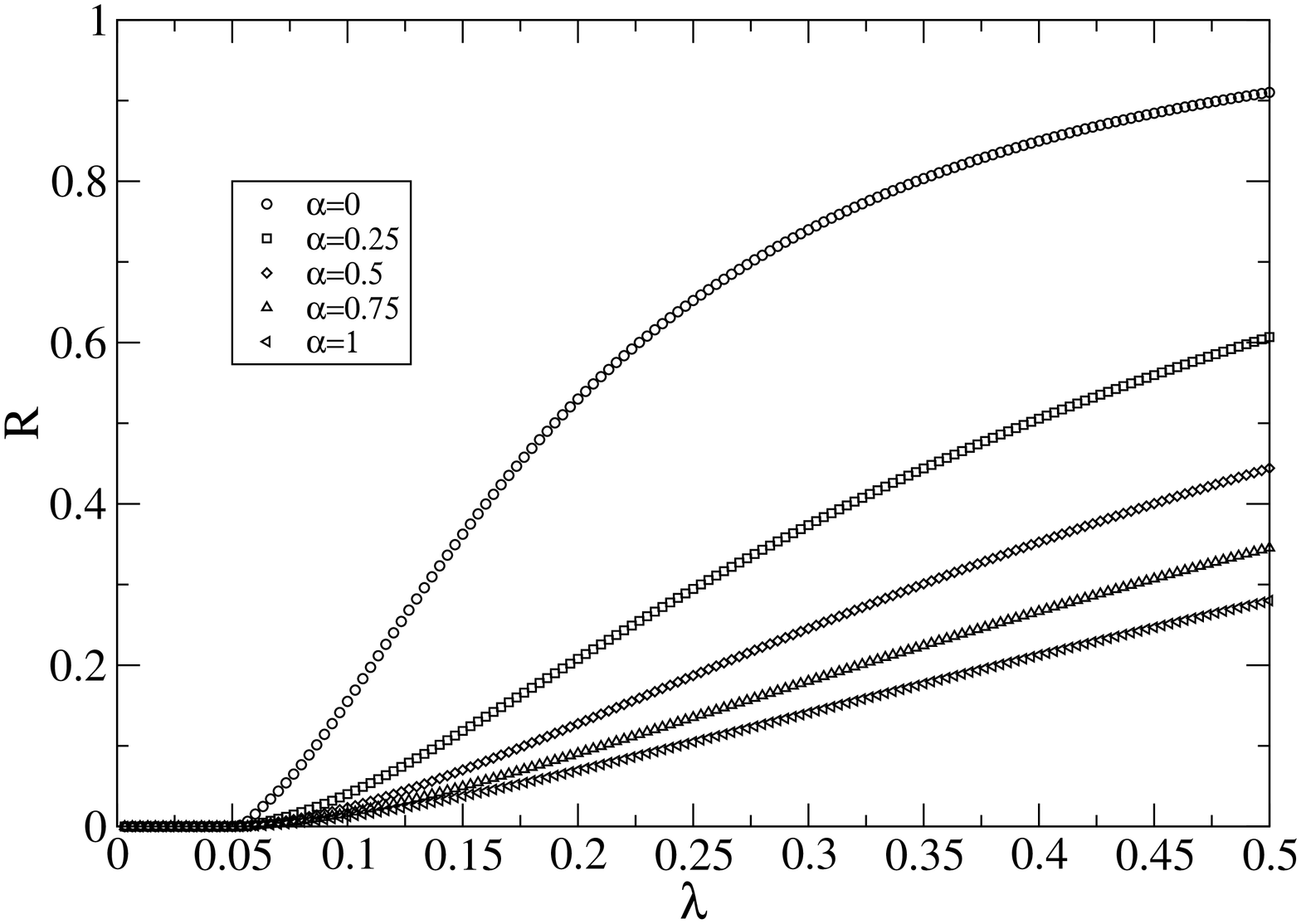,width=2.5in, angle=-90, clip=1}
 \caption{The final size of the rumor, $R$ is shown
as a function of the spreading rate $\lambda$
for the SF network of size $10^6$. The results are shown
for several values of the stifling parameter $\alpha$.}
\end{figure}

\begin{figure}
 \epsfig{file=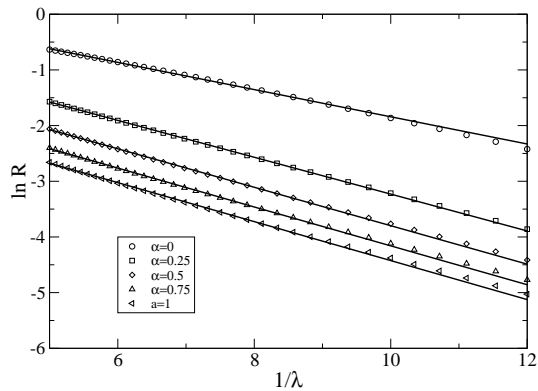,width=2.5in, angle=-90, clip=1}
 \caption{$R$ (in log scale) in the SF network of size $10^6$
   is plotted as a function of $1/\lambda$ and several values of $\alpha$. Solid lines are
   our numerical fits to the stretched exponential form $R= B(\alpha)\exp(-C(\alpha)/\lambda)$.}
\end{figure}

\begin{figure}
 \epsfig{file=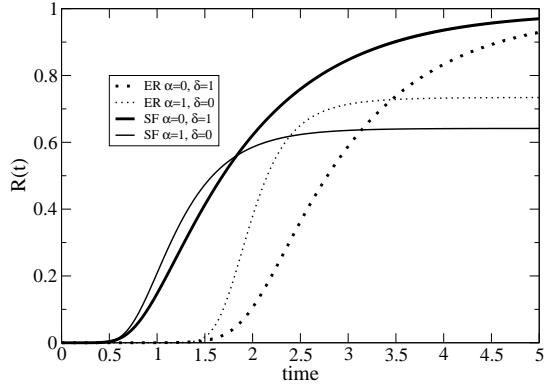,width=2.5in, angle=-90, clip=1} \\
 \caption{Time evolution of the density  of stiflers is shown on
   the ER (dashed lines) and the SF network (solid lines) when the dynamics
   starts with a single spreader node. Results are shown for two sets of values
   of the cessation parameters $\{\alpha=0,\delta=1$\} and $\{\alpha=1,\delta=0\}$. The network sizes
   are $N=10^6$.}
\end{figure}

\begin{figure}
 \epsfig{file=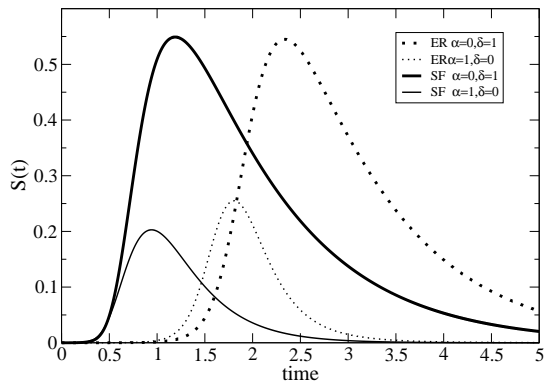,width=2.5in, angle=-90, clip=1} \\
 \caption{Time evolution of the density  of spreaders is shown  for
 the same networks, model parameters and initial conditions as in
 Fig. 5}
\end{figure}

\begin{figure}
 \epsfig{file=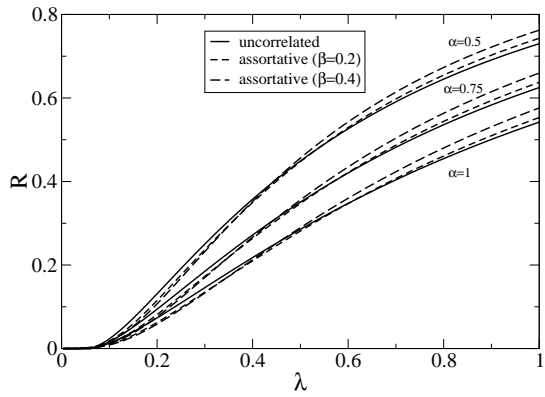,width=2.5in, angle=-90, clip=1} \\
\caption{The final size of the rumor  is plotted as a function of
  $\lambda$ and for several values of $\alpha$ in the SF network of size $10^5$. Results
  are shown in the absence (solid lines) of assortative degree-degree
  correlations and in the presence of such correlations. The correlation
  strengths used are $\beta=0.2$ (short dashed lines) and $\beta=0.4$ (long dashed
  lines).}
\end{figure}

\begin{figure}
 \epsfig{file=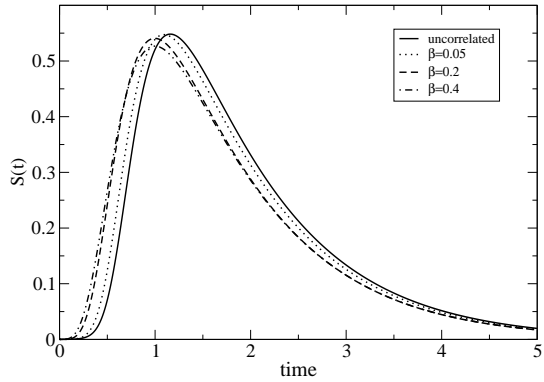,width=2.5in, angle=-90, clip=1} \\
 \epsfig{file=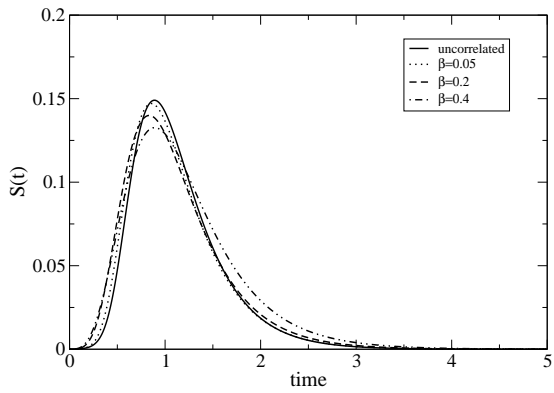,width=2.5in, angle=-90, clip=1} \\
\caption{The impact of assortative correlations on time evolution of
the density  of rumour spreaders.  Results are shown for the SF
networks of size $N=10^5$ and several values of the correlation
strength, $\beta$. The upper panel shows results using $\alpha=0$
and the lower panel those using $\alpha=1$.}
\end{figure}

\end{document}